\documentclass[journal]{IEEEtran}

\usepackage{amssymb}
\usepackage{latexsym}

\usepackage{url}
\usepackage{xcolor}
\usepackage{booktabs}
\usepackage{color,soul}

\usepackage{amsmath}
\usepackage{mathtools}
\usepackage{enumitem}

\usepackage{dblfloatfix}
\usepackage{tikz}       % <---
\usepackage{nicematrix} % <---
\usepackage{framed,multirow}

\usepackage{booktabs,arydshln}
\makeatletter
\def\adl@drawiv#1#2#3{%
        \hskip.5\tabcolsep
        \xleaders#3{#2.5\@tempdimb #1{1}#2.5\@tempdimb}%
    #2\z@ plus1fil minus1fil\relax
        \hskip.5\tabcolsep}
\newcommand{\cdashlinelr}[1]{%
  \noalign{\vskip\aboverulesep
           \global\let\@dashdrawstore\adl@draw
           \global\let\adl@draw\adl@drawiv}
  \cdashline{#1}
  \noalign{\global\let\adl@draw\@dashdrawstore
           \vskip\belowrulesep}}
\makeatother

\usepackage{lipsum}
\usepackage[switch]{lineno}
\usepackage{tcolorbox}
\usepackage[colorlinks=true,linkcolor=red]{hyperref}

\usepackage[labelsep=period,labelfont=bf]{caption}
\usepackage[nameinlink]{cleveref}  % must be after hyperef
\crefname{figure}{Fig.}{\textbf{Figure.}}
\crefname{equation}{Eq.}{\textbf{Eq.}}
\crefname{table}{Table}{\textbf{Table.}}
\crefname{section}{Section}{\textbf{Section}}

\definecolor{newcolor}{rgb}{.8,.349,.1}

\hypersetup{
	colorlinks,
	linkcolor=[rgb]{1.0, 0.0, 0.0},
	citecolor=[rgb]{1.0, 0.2, 0.2},
	urlcolor =[rgb]{0.0, 0.0, 0.8}
}

\begin{document}
% \bstctlcite{IEEEexample:BSTcontrol}

\title{Nuclear Segmentation and Classification: \\ On Color \& Compression Generalization}

\author{Quoc Dang Vu\textsuperscript{*} \and Robert Jewsbury\textsuperscript{*} \and Simon Graham
\and Mostafa Jahanifar \and \\ Shan E Ahmed Raza
\and Fayyaz Minhas \and Abhir Bhalerao \and
Nasir Rajpoot\\ \{quoc-dang.vu, rob.jewsbury, n.m.rajpoot\}@warwick.ac.uk \\ $^*$ Joint first authors, contributed equally

% \institute{
% \textsuperscript{*} Joint First Authors Contributed Equally. \\
% Tissue Image Analytics Centre, University of Warwick\\ \email{\{quoc-dang.vu,rob.jewsbury,n.m.rajpoot\}@warwick.ac.uk}}

% \author{~Quoc~Dang~Vu$^1$, ~Kashif~Rajpoot$^2$, ~Shan~E~Ahmed~Raza$^1$
% and~Nasir~Rajpoot$^{1,*}$\\ $^1$ \{quoc-dang.vu, shan.raza, n.m.rajpoot\}@warwick.ac.uk \\ $^2$ k.m.rajpoot@bham.ac.uk \\ $^*$ Corresponding author
% <-this % stops a space
% \thanks{$^*$Q.D.Vu and R.Jewsbury are joint first authors and contributed equally}
\thanks{Q.D.Vu, R.Jewsbury, S.Graham, M.Jahanifar, S.E.A.Raza, F.Minhas, A.Bhalerao and N.Rajpoot are from the Tissue Image Analytics Centre, Department of Computer Science, University of Warwick, UK}
\thanks{N.Rajpoot is also affiliated with The Alan Turing Institute, London, UK and the Department of Pathology, University Hospitals Coventry \& Warwickshire, UK}
}

% make the title area
\maketitle

\begin{abstract}
Since the introduction of digital and computational pathology as a field, one of the major problems in the clinical application of algorithms has been the struggle to generalize well to examples outside the distribution of the training data. Existing work to address this in both pathology and natural images has focused almost exclusively on classification tasks. We explore and evaluate the robustness of the 7 best performing nuclear segmentation and classification models from the largest computational pathology challenge for this problem to date, the CoNIC challenge. We demonstrate that existing state-of-the-art (SoTA) models are robust towards compression artifacts but suffer substantial performance reduction when subjected to shifts in the color domain. We find that using stain normalization to address the domain shift problem can be detrimental to the model performance. On the other hand, neural style transfer is more consistent in improving test performance when presented with large color variations in the wild.
\end{abstract}


\IEEEpeerreviewmaketitle

\section{Introduction}
The spatial arrangement and morphology of nuclei are important signatures for identifying disease \cite{elston1991NuclearPleomorphism,solis2012LungHistologic}. In cancer, bio-markers such as tumor-infiltrating-lymphocytes \cite{shield2017TILClinical} or Programmed death-ligand 1 (PD-L1) combined positive score for response to immunotherapy \cite{Fusi2015PDL1EA} have been found to be highly correlated with patient survival. In spite of their supposed effectiveness, their adoption and use in clinical settings remain limited due to the high degree of inter and intra observer variation \cite{yamaguchi2018inter} of the derived scores.

To overcome this limitation, many machine learning challenges \cite{kumar2019monuseg,kumar2021Monusac} have been proposed in digital pathology to further the innovation of automating the identification and classification of nuclei instances in Haematoxylin \& Eosin (HE) stained samples. The most recent and perhaps the largest challenge of its kind to date, with 704 participants in 96 teams, the Colon Nuclei Identification and Counting Challenge 2022 (CoNIC) \cite{conic}, was able to identify several powerful deep-learning-based methods for these tasks \cite{pathologyai2022conic,mdc2022conic,stardist2022conic,mbzuai2022conic,arontier2022conic,ciscnet2022conic,denominator2022conic}. These methods achieved competitive performance even on an unseen, large scale cohort. While the results are promising, it is as yet unclear whether these methods are sufficient to achieve generalization across millions of nuclei in the wild for subsequent downstream analyses. 

This question about generalization in the wild is somewhat synonymous with investigating the robustness of models with respect to the shift of input distribution (or domain shift). So far in digital pathology, such investigations have been limited mostly to patch-level classification tasks \cite{stacke2020measuring,yamashita2021learning}. Concurrent to these developments, in natural images, attempts to see how deep neural networks (DNNs) fare against various input alterations have uncovered several important insights. To name a few, existing DNNs have a shape and texture bias \cite{geirhos2018BiasCNN}, this limits the model to reason further about the geometries of their learning targets. Interestingly, this limitation relates to the commonly chosen 3$\times$3 kernel \cite{ding2022BiasLargeKernel}.

With this perspective, it is important to gain better insights about the existing state-of-the art (SoTA) \textit{segmentation-based} methods for identification and classification of nuclear instances. Specifically, in this paper, we investigate how the performance of the top performing approaches in the CoNIC challenge are affected by perturbations to the input. We focus on two forms of domain shift: artifacts introduced by compressing whole slide images and alteration of image colors. We empirically show that:

\begin{figure*}[t]
\centering
\includegraphics[width=1.00\textwidth]{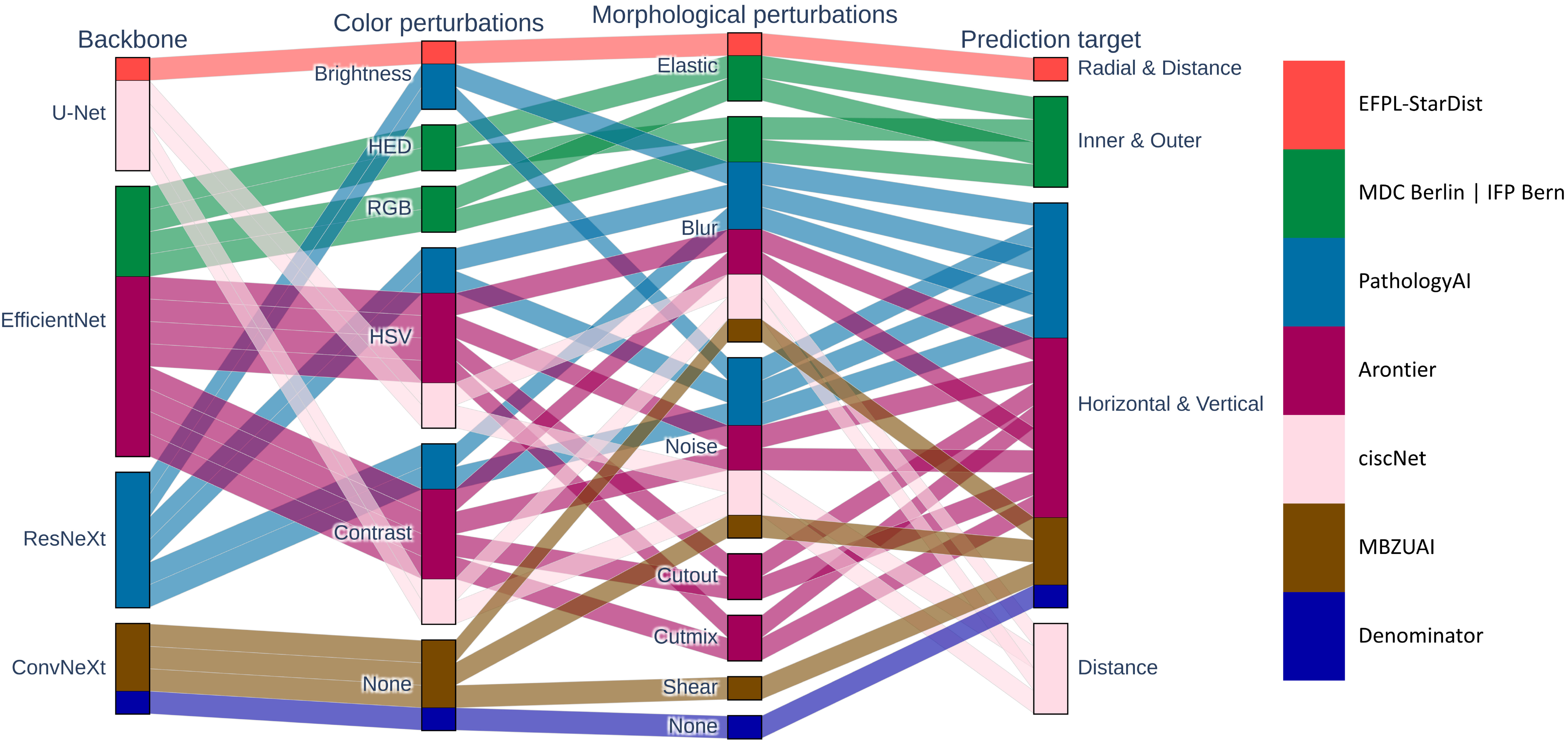}
\caption{Summary of assessed segmentation methods from the CoNIC challenge. Significant aspects including the backbone network family, training augmentations related to color and compression (morphology) and prediction target are shown.
}
\label{f:method_summary}
\end{figure*}
\begin{itemize}
\item Upon the finalization of training, each approach has a virtual center of color distribution that is different from that of the training set;
\item Common augmentations employed during training may not improve robustness against color and morphological perturbations as practitioners might expect;
\item Stain normalization can be more detrimental to performance than beneficial and needs to be employed very carefully.
\end{itemize}

\section{Methodology}
\subsection{Robustness Investigations}

As noted by \cite{hendryck2020robustness}, there is no widely accepted definition for robustness. In this paper, we consider 2 aspects which determine how a sample domain becomes \textit{shifted} compared to that of another sample and investigate how the performance of segmentation-based methods is affected in such cases: a) \textit{Compression related artifacts} and b) \textit{Color variation} between tissue, organs, diseases, scanning devices and staining protocols.

\subsection{Segmentation-based Methods}

We explored the top 7 performing models in the CoNIC challenge that use a segmentation-based approach: Denominator \cite{denominator2022conic},  MBZUAI \cite{mbzuai2022conic},  ciscNet \cite{ciscnet2022conic},  Arontier \cite{arontier2022conic},  PathologyAI \cite{pathologyai2022conic}, MDC Bern - IFP Berlin \cite{mdc2022conic} and EFPL - Stardist \cite{stardist2022conic}. We summarize various aspects of these approaches that can influence how the models fare against the alterations under investigation in \cref{f:method_summary}. All methods follow the encoder-decoder architecture paradigm and only EFPL - Stardist model was trained from scratch.

\subsection{Evaluation Metrics}

The CoNIC challenge utilized a variant of panoptic quality ($PQ$) \cite{kirilov2018panoptic_quality} for multi-class problems called $mPQ^+$ to measure the performance of segmentation and classification of nuclei instances. At IoU$(x_t^a, y_t^a)>a$, the $PQ_t^a$ for each type of nucleus $t$ is defined as
\begin{equation}
\small
\resizebox{.75\linewidth}{!}{$
    \mathcal{PQ}_t^a= 
    \underbrace{\frac{|TP_t^a|}{|TP_t^a|+\frac{1}{2}|FP_t^a|+\frac{1}{2}|FN_t^a|}}_{\text{Detection Quality(DQ)}}
    \times
    \underbrace{{\frac{\sum_{(x_t^a,y_t^a)\in{TP}}{IoU(x_t^a,y_t^a)}}{|TP_t^a|}}}_{\text{Segmentation Quality(SQ)}
    }
$}
\end{equation}
Here, $x_t$ denotes a ground truth GT (GT) instance, $y_t$ denotes a predicted instance, and IoU denotes intersection over union.
% By setting threshold $a\geq0.5$ for IoU$(x^a,y^a)$ or by using Hungarian matching for $a<0.5$ , a unique pairing between a GT and predicted instance is derived.
A unique pairing between a GT and predicted instance is derived when setting the threshold $a\geq0.5$ for IoU$(x^a,y^a)$, or by using Hungarian matching for $a<0.5$. This matching splits all available instances of type $t$ within \textit{an image} into matched pairs (TP), unmatched GT instances (FN) and unmatched predicted instances (FP). In medical images, IoU is a harsh criterion for problems that contain lots of small objects such as nuclei \cite{metrics2022}. However, for downstream analysis, often all predicted nuclei are utilized rather than just a subset being above a certain size. We extend $mPQ^+$ by taking its average across a range $D=\{a \in \mathbb{R}; 0.0\leq a\leq 0.5\}$. This is synonymous to calculating the area under the curve (AUC), thus we define
\begin{equation}
\small
\resizebox{.50\linewidth}{!}{$
    m\mathcal{PQ^+} AUC=\int_{0}^{0.5} \frac{1}{T}\sum_{t}^{T=6}{\mathcal{PQ^+}_t^a} \mathrm{d}{a}
$}
\end{equation}
\noindent To reduce the computation cost, we sampled $\alpha$ with a step of $0.05$ and used the trapezoidal rule to obtain the final results.

\section{Experimental Results}
\subsection{Dataset and Comparison Settings}
\textbf{Training Set}: Participants in the CoNIC challenge utilized the development set of the Lizard dataset \cite{graham2021lizard} for training, validating and selecting their models. In total, there are 431,913 unique nuclei belonging to 6 nuclear categories originating from 5 centers.

\noindent\textbf{Testing Set}: The challenge test set includes images taken from 12 different centers, where 11 of these are completely unseen in the development set. Data from these 11 centers acted as the external test set in the original Lizard publication \cite{graham2021lizard}. The remaining center comprises of additional data extracted from a center that was already considered during training, but from an independent cohort. In total, there exists 103,150 unique nuclei belonging to 6 nuclear categories.

\noindent\textbf{Comparison Settings}: Images from both the Lizard dataset and the testing set were prepared into sets of 256$\times$256 image patches \cite{conic}. To facilitate the discovery for new insights, the selected teams from CoNIC provided a single model that was trained, validated and selected based on a fixed split of the Lizard dataset, provided by the organizers ($80/20$ for training/validation). Analyses in this paper are based on these models rather than the original submissions to the challenge.

\begin{figure*}[t]
\centering
\includegraphics[width=0.90\textwidth]{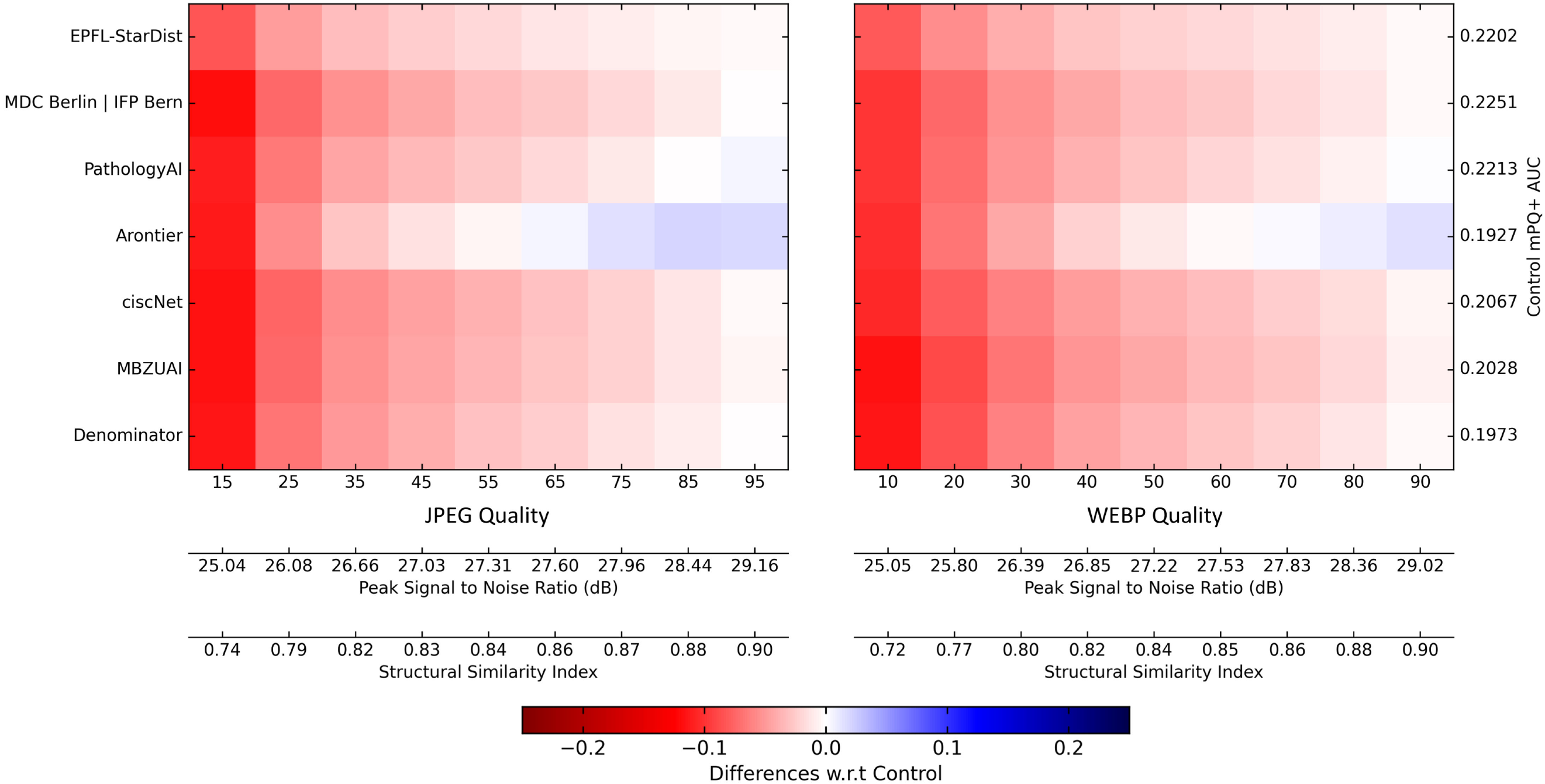}
\caption{Changes in the $m\mathcal{PQ^+} AUC$ of SoTA segmentation-based methods when varying the \textit{quality} parameter in JPEG or WEBP compression.
}
\label{f:compression}
\end{figure*}

\noindent\textbf{Original, Control and Assessed}: We denote \textbf{Original} as the un-altered version of the test set. The whole slide images from which these patches originated could be at different resolutions and compressed by different methods, we aimed to align their information as much as possible before further compressing the data. When shifting the color domain of the test set, we found that neural style transfer\cite{Gatys2015NeuralStyle} (neural stylization) preserves the local characteristics of the images the best at high resolution. We first created a \textbf{Control} version of the test set by up-scaling the \textbf{Original} version to super-resolution via ESRGAN \cite{esrgan} before resizing it back to original resolution. By perturbing this process at various stages, we obtained samples for assessing compression (compressing on \textbf{Control}) and shifting color domains (at super-resolution). We denote these versions of the test set as the \textbf{Assessed}.

\vspace{-5pt}
\subsection{Compression Perturbation}

\noindent We evaluated two major lossy compression methods: JPEG and WEBP. % The former is widely used in computer graphics while the latter is the de-facto choice for transferring and displaying images in current web applications. 
We show how the performance of the selected methods varies across a range of compression qualities in \cref{f:compression}. Intuitively, a majority of the methods experience a decrease in performance the more the images are compressed. Interestingly, the Arontier team's model had noticeable performance gain at some compression levels compared to the \textbf{Control} results. From \cref{f:method_summary}, we hypothesise this could be due to the cutout \cite{tan2019efficientnet} and cutmix \cite{russakovsky2015imagenet} augmentations they employed during training as they would introduce compression-like artifacts. None of the other teams under evaluation employed these augmentation techniques.
% While this is not the only variable compared to other methods we theorise that this is the variable most likely to make their method generalise better with respect to compression based alterations. 

\vspace{-10pt}
\subsection{Color Perturbation}

\begin{figure*}[t]
\centering
\includegraphics[width=0.95\textwidth]{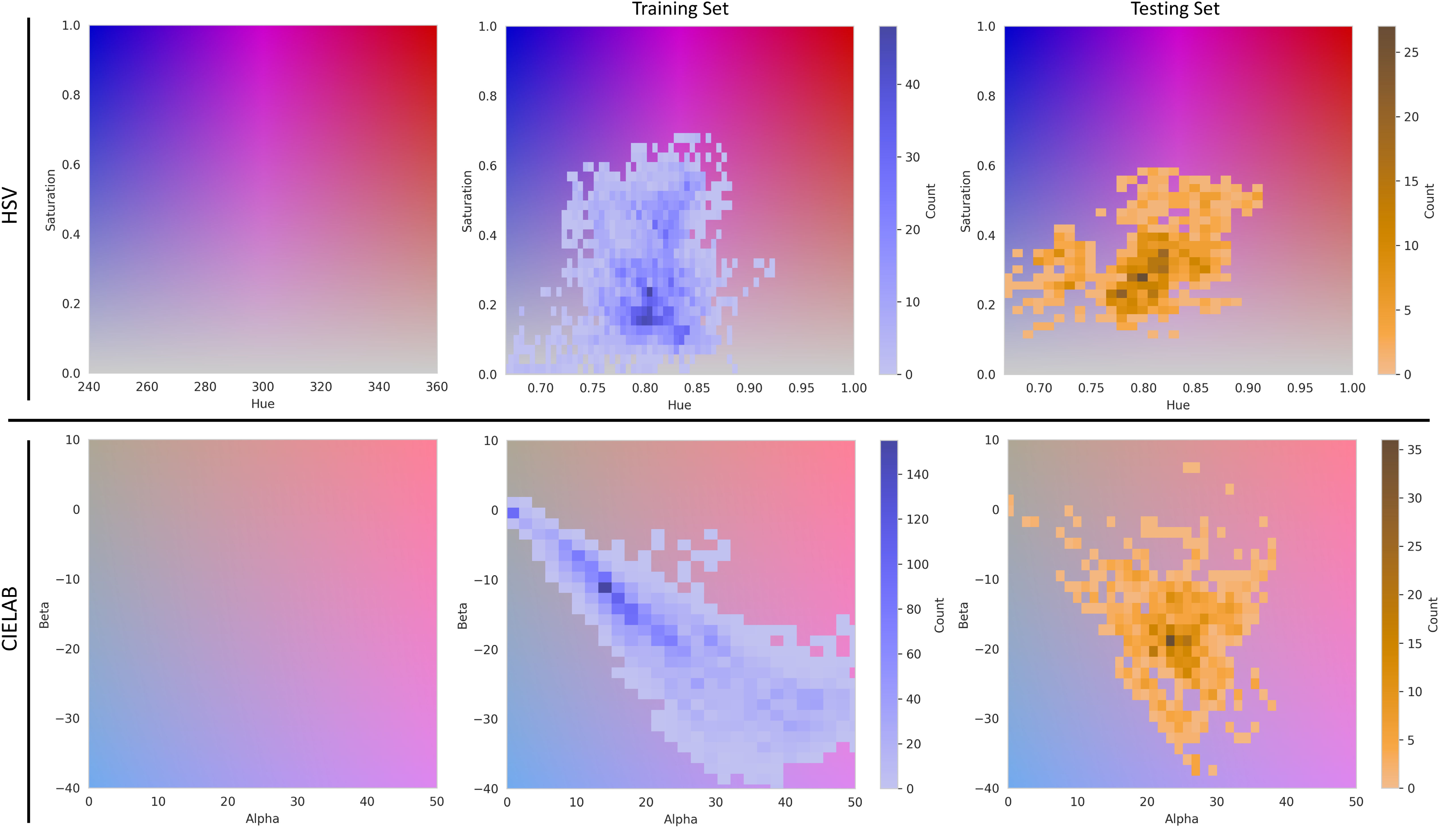}
\caption{Color distribution of image patches within the CoNIC training and testing set using kernel density estimation in HSV and CIELAB color space.
}
\label{f:colorspace}
\end{figure*}

\begin{figure*}[t]
\centering
\includegraphics[width=0.90\textwidth]{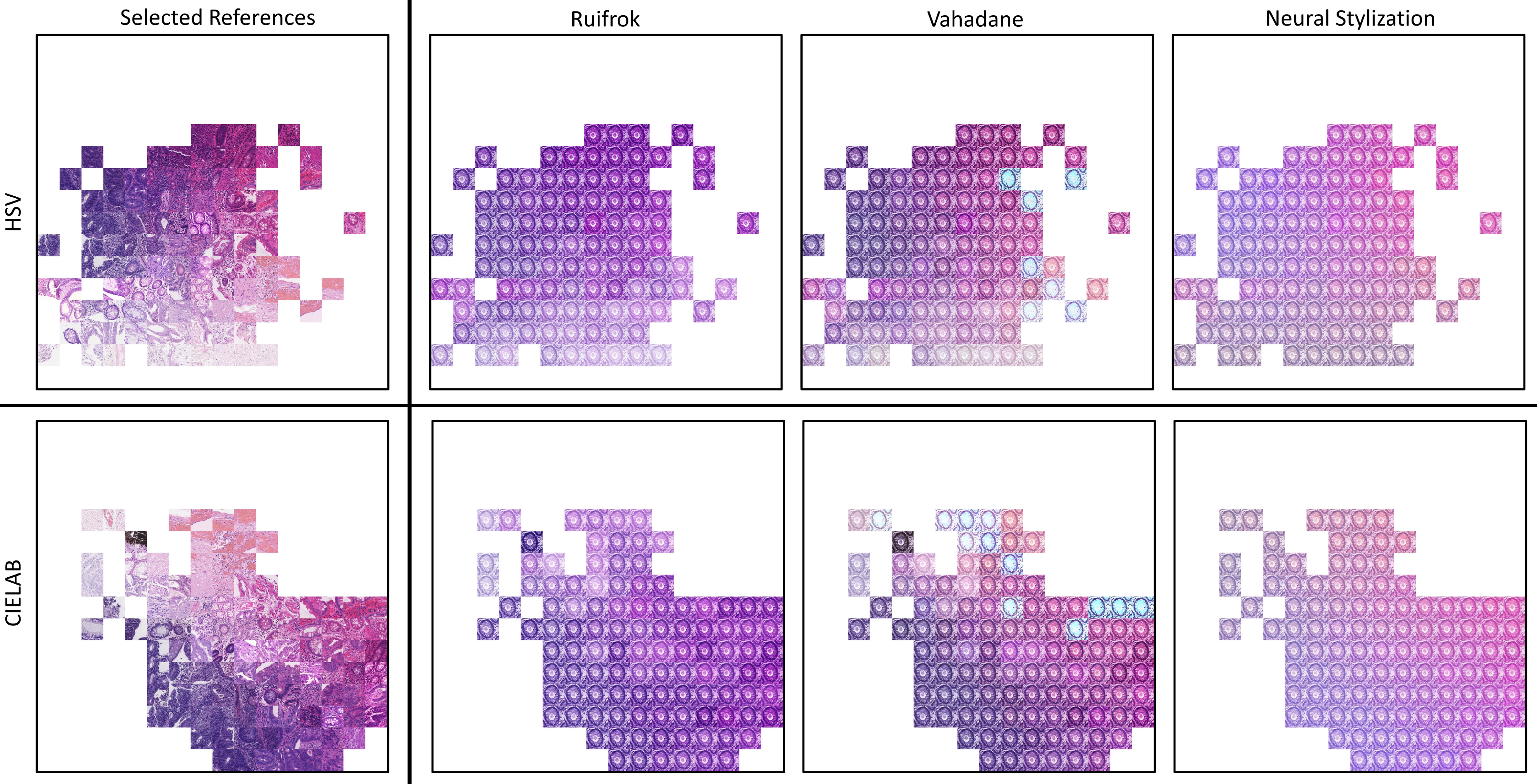}
\caption{Selected reference images within the training set and a test sample that has it color pulled toward the reference at the same location in HSV and CIELAB color space. The color ranges in both color space are the same as in \cref{f:colorspace}.
}
\label{f:references}
\end{figure*}

As each dataset occupies a region within a color space, visualizing how the model performances vary in such a color space is synonymous with approximating the theoretical robustness range of such a model (obtained from a specific training set) with respect to possible samples in the wild. Thus, by identifying how these changes relate to the color distribution of the training and testing set, may provide insights on which aspect of these methods are important for generalization. To ascertain that any such observations are not restricted to a specific color space, we repeat our experiments in both the HSV and CIELAB color spaces.

\noindent\textbf{Color Domains}. We represent the color of each image patch by the mean of its pixel values within that color space. With these values, we establish how the entire training and test set is distributed in terms of Hue ($H$) and Saturation ($S$) for HSV or Alpha ($A$) and Beta ($B$) for CIELAB in  \cref{f:colorspace} via kernel density estimation. Here, we observe that the center of the test set in CIELAB is noticeably different from that of the training set while they heavily overlap in HSV.

\noindent\textbf{Color Sampling and Artificial Domain Shifting:} Having defined the color domains to be investigated qualitatively in \cref{f:colorspace}, further investigations required us shifting the test set toward a color value in such domain while preserving the image compositions. We achieved this by using stain-normalization or by neural stylization. Both techniques however, require a choice of reference image which represents the distribution of the domain we want to shift our target image towards.

To obtain them, we started by simplifying each color space. We restricted the value ranges of the axis within each color space such that they encapsulate the training and testing set distributions. We achieved this by setting $H\in[240, 360]$, $S\in[0.0, 1.0]$ or $A\in[0, 50]$, $B\in[-40, 10]$ based on \cref{f:colorspace}. Afterward, along each axis and within these value ranges, we quantized them into 16 steps, thus creating a total of 256 sampling points (16$\times$16) in HSV or CIELAB respectively. In other words, we obtained 256 samples within the training set which were closest to these 256 colors in each color space. Specifically, by setting the $Value$ ($V$) and $Luminance$ ($L$) to the mean value of the entire training set, we correspondingly obtained unique images within the training set that were the closest to each sampling point. Although there were 256 sampling points, the nearer towards the edge of the training distribution the sampling points became the more likely they shared the same selected images. In such cases, we only retained the ones that have smallest distance within the color space, thus obtaining 101 references for HSV and 108 for CIELAB.

\begin{figure*}[t]
\centering
\includegraphics[width=0.9\textwidth]{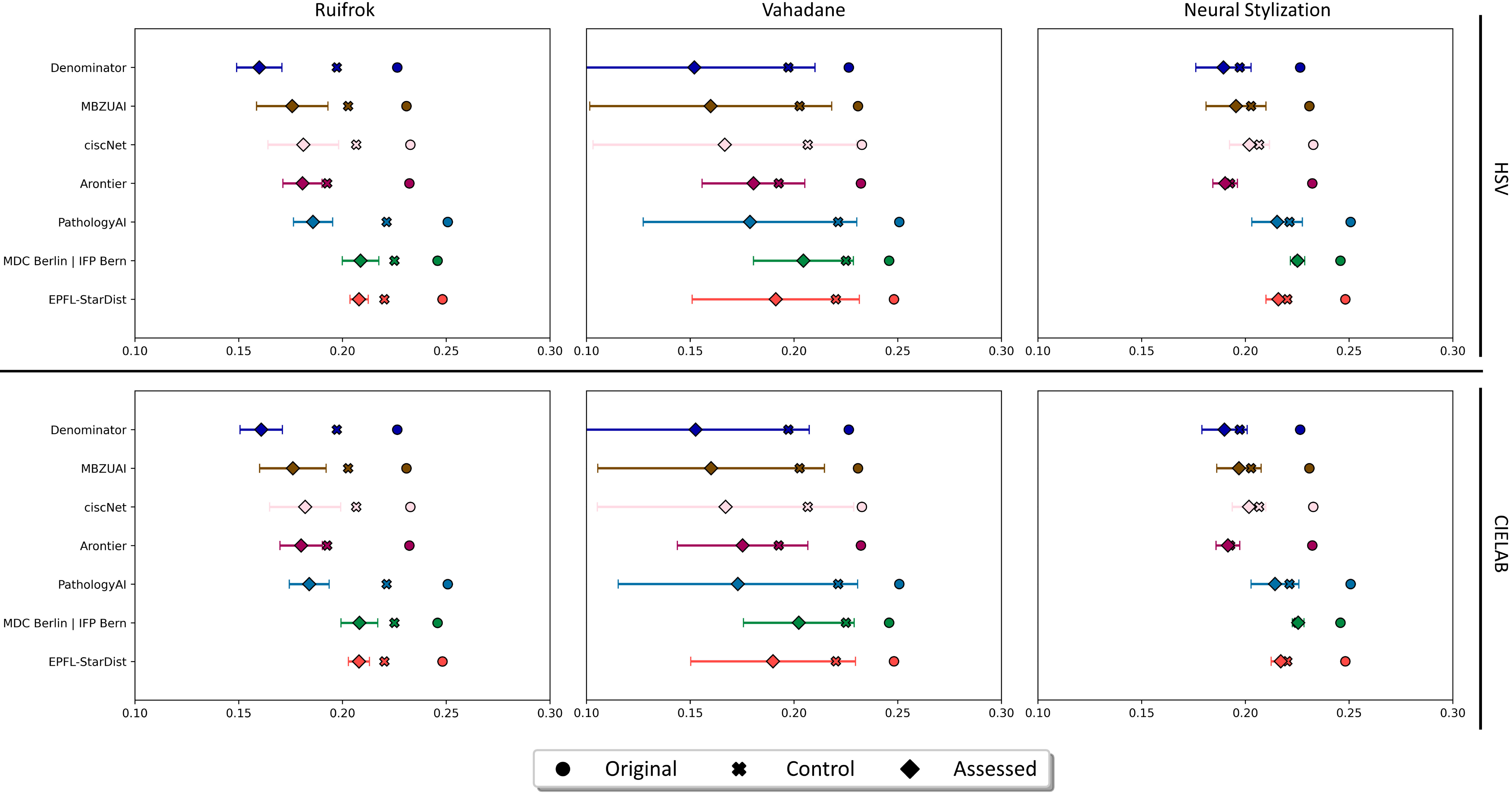}
\caption{
Changes in the $m\mathcal{PQ^+} AUC$ when we artificially shifted the color of the test set using various methods. The box plots are the means and standard deviations.
}
\label{f:variation}
\end{figure*}

Finally, because different color alteration methods behave differently, in order to ensure the validity of the observed phenomena, we also assessed the results obtained by using three different methods: Ruifrok \cite{ruifrok200stainnorm} and Vahadane \cite{vahadane2016structure} stain-normalization and neural style transfer \cite{Gatys2015NeuralStyle}. We provide the sampled references and a standard testing sample with their altered color in \cref{f:references}. Notice that each image in the figure represents a quantized color range as describe above. Additionally, for some reference images, Vahadane did not produce physically realistic samples (the teal colored images). For such cases, we excluded these from our subsequent quantitative analyses detailed in \cref{f:variation}.

\noindent\textbf{Reflected Performances:} For each color alteration method, we compute the average and standard deviation of $mPQ^+ AUC$ across all versions of the testing set in each color space and plot them in \cref{f:variation}. We observed that \textbf{Original} results are higher than that of the \textbf{Control} results. While all methods suffered to varying degrees from the color shift, MDC Bern $|$ IFP Bern and Arontier were affected the least. Across the two color spaces and three color shift techniques, they had the overall smallest standard deviation and reduction in performance. Interestingly, we found stain-normalization to be detrimental to the model performances in general. However, neural stylization was found to be the most stable (having the smallest deviation).

\begin{figure*}[!t]
\centering
\includegraphics[width=1.0\textwidth]{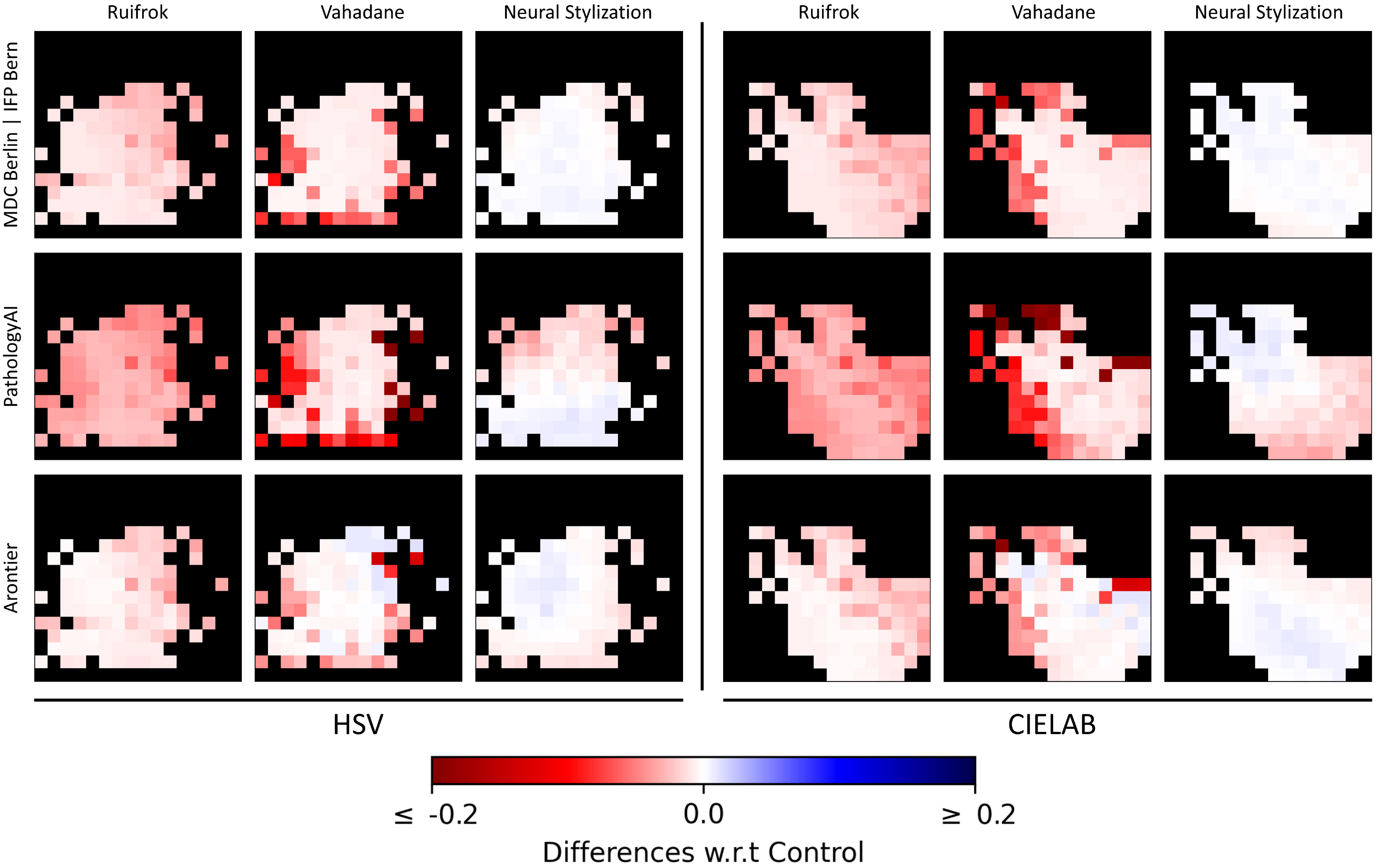}
\caption{
Difference in the $m\mathcal{PQ^+} AUC$ between the \textbf{Control} and the test set where its color was shifted w.r.t each sampled reference in \cref{f:references}. %for 3 of the assessed models. %We only show 3 here due to space constraints.
}
\label{f:variation-sampled}
\end{figure*}

We further show how these performance changes relate to the choice of reference images for the MDC Bern $|$ IFP Bern, Arontier and Pathology AI teams in \cref{f:variation-sampled}. In line with our intuition, moving the color of the testing set toward the center of the training distribution can potentially improve performance, as reflected by PathologyAI. The reference images that demonstrated the smallest reduction in performance (in the case of stain-normalization) or improvement (with neural style-transfer) are found in the center of the training distribution. However, intriguingly, Arontier gained performance when the testing set moved toward areas that are not heavily populated within the training set. Meanwhile, MDC Bern $|$ IFP Bern performed well across the range of sampled color references, with the exception of those being near the training distribution edges. We surmised this is the reason for its large deviation as seen in \cref{f:variation}. Finally, in the context of this task, Vahadane normalization in particular is shown to be highly dependent on the choice of reference image i.e being at the center of the training distribution, regardless of the models.

\section{Conclusions}
We empirically show that SoTA segmentation-based methods are quite robust against compression artifacts. If lossy compression is required, default parameters for JPEG (quality 75) and WEBP (quality 80) are sufficient. However, these models are highly volatile with respect to color variations. We find evidence that stain normalization improves model generalization. Specifically, regardless of the choice of method and color space, we show that stain normalization is more likely to degrade performance than improve it. If altering the color (or stain normalization) of the dataset is required, we suggest using neural style transfer \cite{Gatys2015NeuralStyle} as we found it to provide the most consistent gain in performance. If using stain-normalization, reference images should be from the central area of the training distribution as we found this avoids the outlier regions which resulted in substantial performance loss.

Finally, this work demonstrates that the generalization problem is a concern inspite of the improved performance of computational pathology models in nuclear classification and segmentation tasks achieved in recent years. Our findings highlight the need for further investigations into the mechanism behind these underlying problems.

% \section*{Acknowledgments}

%%Harvard
%% !!!!!! CHANGE THIS BACK FOR SUBMISSION
% \bibliographystyle{model2-names.bst}
% \biboptions{authoryear}
%% !!!!!! CHANGE THIS BACK FOR SUBMISSION
% \newpage
% \newpage
% \newpage
% \newpage
% \newpage
% \newpage
% \newpage
% \newpage
\clearpage

\bibliographystyle{IEEEtran}

\bibliography{reference}

% \newpage
% % a whole empty column
% \vspace*{29.7cm}

% \newpage
% \clearpage
% \appendix
% Reset this table and figure format for Supplementary Material
% \setcounter{subsection}{0}
% \renewcommand{\thesubsection}{A\arabic{subsection}}
% \setcounter{figure}{0}
% \renewcommand{\thefigure}{A\arabic{figure}}
% \setcounter{table}{0}
% \renewcommand{\thetable}{A\arabic{table}}
% \renewcommand{\theequation}{A\arabic{equation}}

\end{document}